\documentclass[preprint,amsmath,amssymb,superscriptaddress,longbibliography,aps,floatfix,footinbib]{revtex4-2}
\usepackage[]{graphicx}
\usepackage{tabularx}
\usepackage[usenames,dvipsnames]{color}
\usepackage{soul}
\usepackage{bm}
\usepackage{amsmath}
\usepackage{lineno}
\usepackage{gensymb}
\usepackage[utf8]{inputenc}

\usepackage{amsfonts}
\usepackage{mathrsfs}
\usepackage{graphicx}
\usepackage{dcolumn}
\usepackage{bm}
\usepackage{color}

\usepackage[colorlinks,bookmarks=false,citecolor=blue,linkcolor=red,urlcolor=blue]{hyperref}
\usepackage{multirow}
\usepackage{physics}
\usepackage{siunitx}
\DeclareSIUnit{\barpressure}{bar}
\DeclareSIUnit\angstrom{\protect \text {Å}}
\usepackage{xcolor}

\begin{document}

\title{Phase-shifted multicomponent spin-charge nematicity in an altermagnet}

\author{Christopher Candelora}
\affiliation{Department of Physics, Boston College, Chestnut Hill, Massachusetts 02467, USA}

\author{Siyu Cheng}
\affiliation{Department of Physics, Boston College, Chestnut Hill, Massachusetts 02467, USA}

\author{Muxian Xu}
\affiliation{Department of Physics, Boston College, Chestnut Hill, Massachusetts 02467, USA}

\author{Keyu Zeng}
\affiliation{Department of Physics, Boston College, Chestnut Hill, Massachusetts 02467, USA}

\author{Hengxin Tan}
\affiliation{School of Physics and Astronomy, Shanghai Jiao Tong University, Shanghai 200240, China}

\author{Younghun Hwang}
\email{younghh@uc.ac.kr}
\affiliation{Electricity and Electronics and Semiconductor Applications, Ulsan College, Ulsan 44610, Republic of Korea}

\author{Binghai Yan}
\affiliation{Department of Physics, Pennsylvania State University: State College, Pennsylvania, USA}

\author{Federico Mazzola}
\affiliation{CNR-SPIN, c/o Complesso di Monte S. Angelo, IT-80126 Napoli, Italy}

\author{Ziqiang Wang}
\email{wangzi@bc.edu}
\affiliation{Department of Physics, Boston College, Chestnut Hill, Massachusetts 02467, USA}

\author{Ilija Zeljkovic}
\email{ilija.zeljkovic@bc.edu}
\affiliation{Department of Physics, Boston College, Chestnut Hill, Massachusetts 02467, USA}

\maketitle

\section{Abstract}
\textbf{Altermagnets host spin-split Fermi surfaces without net magnetization. This intrinsically multicomponent electronic setting raises the possibility that familiar correlated electron phases acquire unconventional spin–charge structure. Here we report the discovery of altermagnetic nematicity in Co$_{0.25}$NbSe$_2$. Using spectroscopic-imaging scanning tunneling microscopy and spin-polarized scanning tunneling microscopy, we find that the three nominally C$_3$-related directions lose rotational equivalence in the zero-field state, in both charge and spin-sensitive tunneling channels. Strikingly, the dominant spin-sensitive component is shifted by one C$_3$ sector relative to the dominant charge component, revealing a phase-shifted spin--charge nematic response. A phenomenological theory shows that altermagnetic order favors a finite relative phase between the charge and spin-sensitive nematic components -- C$_3$ lattice pinning frustrates this preferred offset and selects the observed phase locking. These results establish altermagnetic nematicity as a new form of multicomponent electronic liquid-crystal order and point to a potentially generic route by which altermagnets can transform conventional correlated phases into symmetry-engineered spin--charge orders.}


\section{Introduction}

Correlated electronic phases in quantum solids, including superconductivity, density waves and electronic nematicity, are shaped not only by interactions, but also by the symmetries of the electronic states from which they emerge. Altermagnets, which combine magnetic compensation with spin-split electronic bands to produce spin-polarized systems without net magnetization  \cite{Hayami2019, Smejkal2020, Yuan2020, Mazin2021, Smejkal2022_1, Smejkal2022_2,Krempasky2024, Jiang2025sdw, Reimers_2024} have introduced a new setting for this principle. Emerging theoretical work suggests that conventional correlated phases can acquire unusual forms when embedded in this symmetry-structured spin-polarized background. For instance, in superconducting hybrids, altermagnetic spin splitting can generate finite-momentum Cooper pairing and impose strong constraints on allowed pairing symmetries \cite{Zhang2024FiniteMomentum,Chakraborty2024ZeroField,Chakraborty2025Constraints}. Theory has also proposed spin-, charge-, and pair-density-wave instabilities in altermagnetic metals, where momentum-dependent spin polarization can produce wavevector structures distinct from those in ordinary metals \cite{Parthenios2025SpinPair,Yang2026SpinResolvedCDW,Ding2025AnomalousCDW, Zou2026Superconducting}. These developments raise an experimentally open question: can a conventional correlated order be transformed into a qualitatively new spin–charge phase by altermagnetic symmetry?

Electronic nematicity, in which electronic degrees of freedom lower crystalline rotational symmetry by selecting a preferred direction, provides a particularly direct test of this question because it acts on the same spatial symmetries that organize altermagnetic spin polarization in many altermagnets \cite{Smejkal2022_1, Smejkal2022_2, Krempasky2024, jungwirth_symmetry_2026, Leeb2024, Zhou2024}. Such anisotropic electronic states have been extensively explored in correlated metals and unconventional superconductors \cite{Fradkin2010nematic,Lawler2010,Chuang2010nematic, Chu2012nematic,Rosenthal2014nematic, Singh2015FTSnematic,Yi2017FeSCsreview,Nie2022Charge-density-wave-drivenSuperconductor, Li2023ElectronicMetal,Li2022RotationKV3Sb5, Jiang2024Sc166nematic}. In these familiar settings, even when charge, orbital, lattice or spin responses are intertwined, they are typically organized by a single nematic director. Altermagnets provide a different situation: their compensated spin-split Fermi surfaces form an intrinsically multicomponent electronic background that is wavevector dependent, so different nematic components can, in principle, manifest in unexpected ways. 

Intercalated transition metal dichalcogenides have recently emerged as a versatile platform for realizing altermagnetism, where the choice of intercalated magnetic species and the host transition metal can stabilize different magnetic ground states \cite{sah2025altermagnetismkagomeflatband, dayroberts2026altermagneticmaterialslibraryintercalated, Hatanaka2023MagneticIntercalations, Regmi2024, mandujano2025, naik_evolution_2022, parkin_3_1980, voorhoeve-van_den_berg_low-temperature_1971, xie_structure_2022, morosan_sharp_2007, mangelsen_large_2020, park_composition_2024, wu_highly_2022, kousaka_emergence_2022, gubkin_crystal_2016, rodriguez_magnetic-crystallographic_2011, Ghimire2018}. Among this family, cobalt-intercalated NbSe$_2$, Co$_{0.25}$NbSe$_{2}$, has attracted particular attention \cite{Regmi2024, candelora_discovery_2026, deVita_switch, Dale_2026, sakhya2025electronicstructurelayeredaltermagnetic}, combining a triangular lattice geometry with an ordered 2 $\times$ 2 Co superstructure (Fig.~\ref{fig:1}a,b). Angle-resolved photoemission spectroscopy and density functional theory calculations have revealed robust spin splitting of the electronic bands consistent with altermagnetic behavior in this system (Fig.~\ref{fig:1}c) \cite{deVita_switch, Dale_2026, sakhya2025electronicstructurelayeredaltermagnetic}. 

In this work, we investigate the zero-field electronic symmetry of Co$_{0.25}$NbSe$_{2}$ using spectroscopic-imaging STM and spin-polarized STM. By comparing the amplitudes of the three 2$a_0$ modulation peaks that should be nominally equivalent, we find that the underlying C$_3$ rotational symmetry of the electronic structure is broken. The inequivalence of these three directions is observed consistently in both the charge and spin channels, revealing an electronic nematic instability embedded within the altermagnetic state. Interestingly, the dominant projected spin-resolved component is rotated relative to the dominant charge component, giving rise to a phase-shifted spin--charge nematic state. Our phenomenological model attributes this offset locking to coupling between nematicity and the altermagnetic order parameter. Together, these results show that altermagnetism can reshape electronic nematicity into a multicomponent spin--charge order, providing a route to symmetry-engineered correlated phases in spin compensated magnetic metals.

\section{Results}

\subsection{Zero-field rotational symmetry breaking in the charge channel}

\begin{figure}
    \centering
    \includegraphics[width = \textwidth]{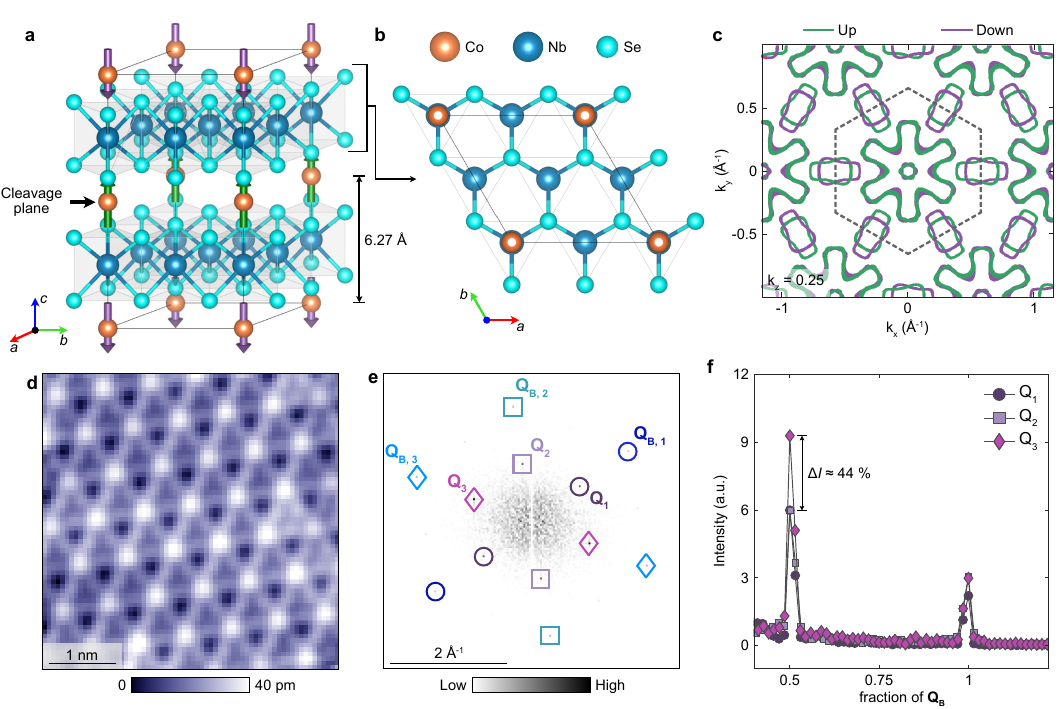}
    \renewcommand{\baselinestretch}{1}
    \caption{\textbf{Crystallographic and electronic structure of Co$_{0.25}$NbSe$_2$}. \textbf{a}, Three dimensional unit cell crystal structure of Co$_{0.25}$NbSe$_2$. \textbf{b}, Atomic structure in the $ab$-plane. \textbf{c}, DFT spin-polarized 2D Fermi surface at $k_z$ = $\pi$/2$c$ with spin-up bands in green and spin-down bands in purple, recreated from Candelora, et al. \cite{candelora_discovery_2026}. \textbf{d}, Representative STM topograph of the Se-terminated surface. \textbf{e}, Representative Fourier transform of the Se-terminated surface. The 2$a_0$ modulations due to the underlying Co are denoted by \textbf{Q}$_i$ ($i$ = 1, 2, 3), with atomic Bragg peaks marked by \textbf{Q}$_{B, i}$ ($i$ = 1, 2, 3). (\textbf{d}) is a smaller region within the area used to obtain the Fourier transform in (\textbf{e}). \textbf{f}, Intensity linecuts along each $\textbf{q}$ direction in Fourier space, showing an intensity difference between \textbf{Q}$_3$ and the other two directions (\textbf{Q}$_1$ and \textbf{Q}$_2$) at zero magnetic field, while all three Bragg peaks maintain almost the same intensity. STM setup conditions: (\textbf{d}) $V_{sample}$ = 50 mV, $I_{set}$ = 200 pA. The magnetic field is set to 0 T.}
    \label{fig:1}
\end{figure}

We cleave bulk single crystals of Co$_{0.25}$NbSe$_{2}$ in ultra-high vacuum at cryogenic temperature and immediately insert them into the microscope head (Methods). The crystals have a 2D layered structure, with the 2 $\times$ 2 superstructure of Co atoms residing between adjacent Se-Nb-Se slabs (Fig.~\ref{fig:1}a). Previous scanning tunneling microscopy (STM) experiments have provided real-space evidence for the $2 \times 2$ Co ordering within the van der Waals gap, which doubles the original unit cell of pristine NbSe$_2$ into a 2 $\times$ 2 super-cell (Fig.~\ref{fig:1}d,e) \cite{candelora_discovery_2026}. While that work established the presence and magnetic-field tunability of the $2a_0$ modulation, it left open whether, in the zero-field state, the three nominally C$_3$-related modulation directions are electronically equivalent. 

\begin{figure}
    \centering
    \includegraphics[width = \textwidth]{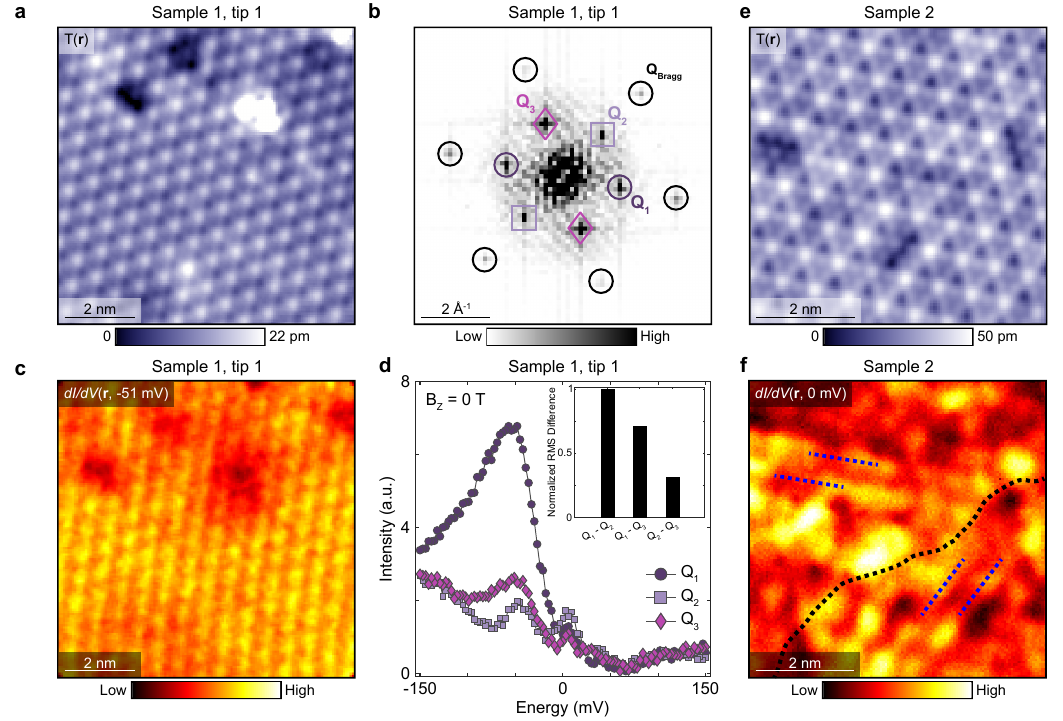}
    \renewcommand{\baselinestretch}{1}
    \caption{\textbf{Rotational symmetry breaking in the charge channel.} \textbf{a, b}, STM topograph of the Se-terminated surface on sample 1 (\textbf{a}) and its corresponding Fourier transform (\textbf{b}). \textbf{c} $dI$/$dV$ map over the same area in (\textbf{a}) which was box filtered for visual clarity. \textbf{d}, Intensity of the 2$a_0$ peaks as a function of energy taken from the Fourier transform of the unfiltered, drift-corrected d$I$/d$V$ maps over the area in (\textbf{a}). The inset in \textbf{d} shows the normalized root mean square (RMS) difference between each peak, with a higher RMS signaling a larger difference. This was normalized by setting the largest RMS to 1. \textbf{e}, STM topograph of the Se-termination on a different sample with a different tip. \textbf{f}, $dI$/$dV$ map over the same area shown in (\textbf{e}) with blue dashed lines marking unidirectional behavior and the black dotted line marking the domain boundary. STM setup conditions: (\textbf{a}) $V_{sample}$ = 150 mV, $I_{set}$ = 100 pA; (\textbf{c}) $V_{sample}$ = 150 mV, $V_{exc}$ = 3 mV (RMS), $I_{set}$ = 100 pA; (\textbf{e}) $V_{sample}$ = 50 mV, $I_{set}$ = 150 pA; (\textbf{f}) $V_{sample}$ = 50 mV, $V_{exc}$ = 5 mV (RMS), $I_{set}$ = 150 pA. The magnetic field is set to 0 T.}
    \label{fig:2}
\end{figure}

Fourier transforms of STM topographs of the Se surface acquired with non-spin-polarized STM tips already show an inequivalence of the 2$a_0$ modulation peaks -- for example, $\mathbf{Q}_{2a0}^3$ is substantially higher in amplitude than $\mathbf{Q}_{2a0}^2$, despite the corresponding atomic Bragg peaks being nearly identical (Fig.~\ref{fig:1}f). We examine this inequivalence in greater depth using Fourier transforms of d$I$/d$V$(\textbf{r}, $V$) maps. In a C$_3$-rotationally symmetric state, the three peaks $\mathbf{Q}_{2a0}^i$ would be equivalent. Instead, we again find one direction to be notably stronger compared to the other two that appear more similar, as observed directly in Fourier space and quantified via root-mean-square analysis (Fig.~\ref{fig:2}a-d). The observation of domains with anisotropy along different directions within the same field-of-view and with the same tip is inconsistent with a purely tip-induced anisotropy and instead supports an intrinsic symmetry-broken electronic state (Fig.~\ref{fig:2}e,f). Since this symmetry breaking lowers rotational equivalence without introducing an additional translational periodicity beyond the existing 2 $\times$ 2 Co superstructure, we refer to it as an electronic nematic state.

\subsection{Spin-resolved rotational symmetry breaking}

\begin{figure}
    \centering
    \includegraphics[width = \textwidth]{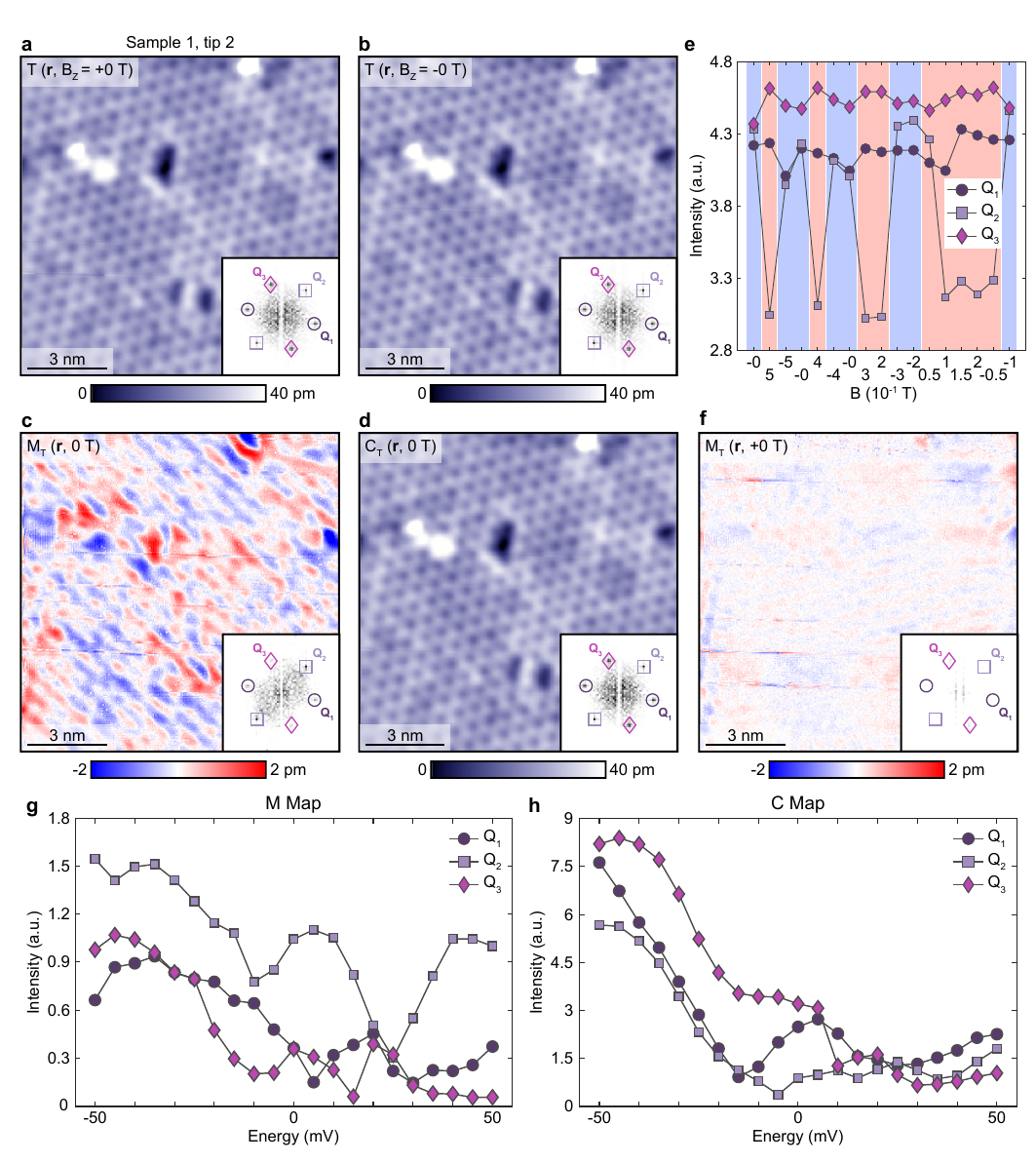}
    \renewcommand{\baselinestretch}{1}
    \caption{\textbf{Spin-polarized STM imaging to disentangle spin and charge induced symmetry breaking}. \textbf{a}, STM topograph of the Se termination with a spin-up polarized STM tip. Inset to (\textbf{a}) is the Fourier transform of the area with the 2$a_0$ peaks marked for the three directions. \textbf{b}, STM topograph over the same area as in (\textbf{a}) but with spin-down polarized tip, and the corresponding Fourier transform (inset). \textbf{c}, Magnetic contrast map ($M_T$(\textbf{r})) obtained by subtracting the STM topographs in (\textbf{a,b}), with the corresponding Fourier transform (inset). \textbf{d}, Charge contrast map ($C_T$(\textbf{r}) obtained by adding the topographs in (\textbf{a,b}), and corresponding Fourier transform (inset).} 
    \label{fig:3}
\end{figure}

We investigate whether the observed symmetry breaking extends to the spin channel using spin-polarized STM. Figures~\ref{fig:3}a,b show topographs acquired over the same region with opposite tip spin polarization. While they appear nearly identical by eye, their difference isolates the spin-dependent contribution. By aligning and subtracting the two images, we obtain a spin-resolved map $M_T(\mathbf{r})$ (Fig.~\ref{fig:3}c), which reveals modulations strongly enhanced along a single lattice direction. Conversely, the sum of the two images yields a spin-averaged map $C_T(\mathbf{r})$ (Fig.~\ref{fig:3}d), in which the out-of-phase spin contributions cancel, leaving the dominant charge signal. To verify that the extracted spin contrast is not an artifact of the subtraction procedure, we repeat the analysis on two topographs acquired with the same tip spin polarization, obtained by reversing the tip spin and then reversing it back. Their difference yields a negligible signal (Fig.~\ref{fig:3}f), confirming that $M_T(\mathbf{r})$ isolates a genuine spin-dependent contribution. This procedure is highly reproducible: upon reversing the tip spin polarization, we observe a systematic and reversible redistribution of spectral weight among the $\mathbf{Q}_i$ peaks, with the most pronounced changes occurring at $\mathbf{Q}_2$ (Fig.~\ref{fig:3}e).

To disentangle the spin and charge contributions as a function of energy, we extend this analysis to spectroscopic maps. We define the spin-resolved and spin-averaged signals as $M(\mathbf{r}, V) = [dI/dV^{\uparrow}(\mathbf{r}, V) - dI/dV^{\downarrow}(\mathbf{r}, V)]/2$ and $C(\mathbf{r}, V) = [dI/dV^{\uparrow}(\mathbf{r}, V) + dI/dV^{\downarrow}(\mathbf{r}, V)]/2$, respectively. The spin-resolved $M(\mathbf{r}, V)$ maps exhibit a pronounced redistribution of spectral weight among the three wavevectors, with one peak strongly enhanced and the relative intensities evolving markedly with energy. Most notably, the $\mathbf{Q}_2$ peak, which is comparatively weak in the spin-averaged channel over much of the measured energy range, gives the largest projected spin-resolved contrast for the measured tip polarization (Fig.~\ref{fig:3}g,h, Extended Data Fig.~\ref{si:3}a-b). The distinct and energy-dependent behavior of the $\mathbf{Q}_{2a_0}^i$ peaks therefore demonstrates that the rotational symmetry breaking is reflected in the spin-sensitive tunneling response, pointing towards a spin--charge nematic state. Moreover, the difference between the spin-averaged and spin-resolved channels shows that the spin-polarized local density of states associated with the three $2a_0$ wavevectors is not simply proportional to the charge modulation response. 

\subsection{Local defect response in the symmetry-broken phase}

To further examine the local electronic response within the symmetry-broken phase, we study recurring point-like defects observed in the same field of view. We focus on a class of defects that appear at the same intra-unit-cell registry, with two examples outlined in Fig.~\ref{fig:4}a,b. Based on their position relative to the surrounding Se lattice, these defects are centered between three neighboring Se atoms (Fig.~\ref{fig:4}e), a local environment that is nominally structurally symmetric under C$_3$ rotations. In $dI/dV(\mathbf{r},V)$ maps, these defects generate three-lobed local density-of-states patterns aligned with the three in-plane lattice directions (Fig.~\ref{fig:4}f-k). The lobe intensities are generally unequal and evolve with bias (Extended Data Fig.~\ref{si:1}q-w, Extended Data Fig.~\ref{si:2}). Such three-lobed defect patterns are a natural local response to the observed nematic background. 

Importantly, the hierarchy of the three lobes is not identical for all defects. Within the same field of view and under identical measurement conditions, different defects exhibit different local lobe patterns (Fig.~\ref{fig:4}f-k, Extended Data Fig.~\ref{si:1}c-p), while the surrounding $2a_0$ spin modulation remains dominated by a single lattice direction. This contrast is difficult to reconcile with a fixed anisotropic tip imposing a uniform directional preference across the image. Instead, the defect-centered local density-of-states patterns provide a local check that the observed anisotropy of the $2a_0$ modulation is not generated by tip anisotropy. Together with the charge-channel domain structure and the spin-resolved measurements, these observations support an intrinsic electronic origin of the rotational symmetry breaking.

\subsection{Phenomenological model of phase-shifted spin--charge nematicity}

The observed rotation between the charge and spin-resolved nematic responses can be captured by a minimal Landau theory for a $C_3$-symmetric altermagnetic metal (Methods). We describe the two spin-split Fermi surfaces by nematic distortions $N_+$ and $N_-$ (for instance, green and purple colored bands in Fig.~\ref{fig:1}c), and form their symmetric and antisymmetric combinations, $\Phi=(N_+ + N_-)/2=|\Phi|e^{i\varphi_{\Phi}}$ and $\Psi=(N_+ - N_-)/2=|\Psi|e^{i\varphi_{\Psi}}$. The field $\Phi$ represents the charge-like nematic component measured in the spin-averaged tunneling channel, while $\Psi$ represents the spin-sensitive nematic component measured by spin-polarized STM through its projection onto the tip polarization. In the altermagnetic state, the order parameter $\Delta_A$ that produces spin-split Fermi surfaces allows an imaginary bilinear coupling between $N_+$ and $N_-$, which becomes a locking term of the form
\[
F_{\rm lock} = -4J_I \Delta_A|\Phi||\Psi|\sin(\varphi_\Psi-\varphi_\Phi),
\]
where $\phi_\Phi$ and $\phi_\Psi$ are the phases of the charge-like and spin-sensitive nematic components. This term strongly disfavors $\varphi_\Psi=\varphi_\Phi$, and it is minimized when there is a $\pi/2$ phase difference between $\Phi$ and $\Psi$. However, this preferred offset is incompatible with the discrete $C_3$ Potts angles available to the nematic components. The lattice therefore frustrates the altermagnetic locking tendency and selects the nearest neighboring Potts sector, producing the phase-shifted spin--charge nematic state in which the dominant projected spin-resolved modulation direction is offset from the dominant charge direction (Fig.~\ref{fig:5}), as observed experimentally. We note that this shift is robust regardless of the sign of $J_I\Delta_A$, which only determines whether $\Phi$ is rotated clockwise or counterclockwise relative to $\Psi$. Within this phenomenological description, the observed state is an altermagnetic nematic state in which nematic distortions of two spin-split C$_3$-symmetric Fermi surfaces generate intertwined charge and spin nematicity occupying neighboring Potts sectors.

\subsection{Conclusion and discussion}
Using STM, spin-polarized STM and theory, we discover a multicomponent phase-shifted spin–charge nematic state, occurring in altermagnet Co$_{0.25}$NbSe$_2$. The Co-induced 2 $\times$ 2 superstructure provides three nominally equivalnet C$_3$-related channels; in the spin-averaged tunneling response, these channels lose rotational equivalence and one direction is selected, defining a charge-like nematic component. Spin-polarized STM shows that the corresponding spin-sensitive response is also nematic, but with a different dominant wavevector. Thus, the spin-sensitive nematic component is not simply a passive copy of the charge modulation. Instead, it is rotated by one C$_3$ sector relative to the charge component. This phase offset is the defining feature of the observed state: the charge and projected spin-resolved responses break the same rotational symmetry, but they do not select the same electronic direction. This state can be naturally viewed as a multicomponent nematic: the primary nematic distortions occur on the two spin-split altermagnetic Fermi surfaces, while the experimentally observed charge-like and spin-sensitive responses correspond to their symmetric and antisymmetric combinations.

Our phenomenological model provides a natural interpretation of this offset locking. In the altermagnetic state, the charge-like and spin-sensitive nematic components can couple through the altermagnetic order parameter. This coupling favors a $\pi/2$ relative phase between the two components, while the C$_3$ lattice restricts each component to discrete Potts-like orientations. The altermagnetic locking tendency and the crystalline clock anisotropy cannot be simultaneously optimized, so the spin-sensitive component is displaced into a neighboring Potts sector relative to the charge component. In this sense, the altermagnetic background does not merely coexist with nematicity -- it changes how the charge and spin-sensitive nematic components lock to one another.

Angle-resolved photoemission spectroscopy on samples from the same growth batch reveals spin-split band structure expected for the altermagnetic state \cite{deVita_switch}, confirming that the STM-observed symmetry breaking occurs in a compensated spin-polarized metal. This raises intriguing questions to be explored in future experiments: how nematic domains modify momentum-dependent spin splitting, whether the spin-polarized spectral weight can be redistributed among symmetry-related wavevectors, how such symmetry lowering appears in anisotropic spin and charge transport, and how the nematicity onsets and evolves with temperature. Co$_{0.25}$NbSe$_2$ therefore provides a platform for exploring how altermagnetism reshapes electronic nematicity into symmetry-engineered spin--charge order.

It is important to note that the spin-resolved channel measures the projection of the sample spin-polarized local density of states onto the tip polarization. If the three 2$a_0$ modulations were associated with a simple collinear spin configuration carrying the same spin direction and equivalent spin-polarized spectral weight, reversing the tip polarization would affect \textit{all} three $\mathbf{Q}_i$ components in the same way, in contrast to what we observe experimentally. Similarly, a uniformly canted configuration would preserve a common projection factor for all three directions, and therefore could not produce the observed wavevector-dependent redistribution of spin contrast. The observed response, including the large projected $\mathbf{Q}_2$ contrast for the measured tip polarization despite its weaker charge-channel intensity, instead requires the spin-polarized components to be inequivalent. Given the out-of-plane magnetic structure of Co$_{0.25}$NbSe$_2$, the most natural interpretation is that the three 2$a_0$ components carry inequivalent out-of-plane spin-polarized spectral weights.

Several internal checks distinguish this electronic anisotropy from instrumental effects. A spin-polarized tip can influence the measured magnetic contrast through projection, but such a projection does not account for the corresponding anisotropy in the spin-summed channel. A fixed tip-shape anisotropy would also be expected to impose a uniform directional preference over a field of view, whereas the charge maps show regions with different preferred modulation directions. Finally, defect-centered local density-of-states patterns measured under identical conditions display different local lobe hierarchies while the surrounding $2a_0$ modulation remains dominated by a single direction. Taken together, these observations support an intrinsic origin of the rotational symmetry breaking.

The broader significance of these results goes beyond the realization of nematicity in an altermagnet. Nematicity has long been studied in correlated electron systems such as cuprates, Fe-based superconductors, heavy-fermion systems and kagome metals \cite{Fradkin2010nematic,Lawler2010,Chuang2010nematic, Chu2012nematic,Rosenthal2014nematic, Singh2015FTSnematic,Yi2017FeSCsreview, Okazaki2011Rotationalsub2/sub,Nie2022Charge-density-wave-drivenSuperconductor, Li2023ElectronicMetal,Li2022RotationKV3Sb5, Jiang2024Sc166nematic}, and recent work has revealed three-state nematicity tied directly to magnetic degrees of freedom in certain van der Waals antiferromagnets \cite{Yao2025FePSe3,Kirstein2026NatComm}. What distinguishes Co$_{0.25}$NbSe$_2$ is that nematicity emerges in an altermagnetic metal, where compensated magnetic order produces spin-polarized electronic structure without net magnetization. In this setting, nematicity does not appear as a conventional single-director anisotropy. Instead, enabled by the altermagnetic parent state, it becomes a multicomponent spin--charge order, showing that a familiar electronic liquid-crystal phase can acquire a qualitatively different form inside an altermagnetic background. Thus our results establish altermagnetic nematicity as a fundamentally distinct form of correlated electronic order, distinguished by the intertwining of spin-split Fermi surface geometry, lattice Potts anisotropy, and  phase-shifted spin-charge nematic order. More broadly, our work points to a symmetry principle for altermagnetic quantum matter: correlated orders can inherit the multicomponent structure of compensated spin-split Fermi surfaces, and be reshaped into electronic phases not readily available in other systems.\\

\noindent{\bf Author contributions}\\ 
C.C. performed STM measurements with the help from S.C. and M.X. Z.W. and K.Z. provided the theoretical model. Y.H. synthesized the bulk single crystals. F.M. provided ARPES measurements. H.T. and B.Y. performed DFT calculations. I.Z., C.C. and Z.W. wrote the paper with the input from all the authors. I.Z. supervised the project. \\
\\
\noindent{\bf Methods}\\
\indent \textit{\textbf{Sample growth:}} Synthesis of Co$_{0.25}$NbSe$_2$ single crystals was performed using the chemical vapor transport (CVT) technique. Starting materials consisted of high-purity powders of cobalt (Co, 99.99 \%), niobium (Nb, 99.999 \%), and selenium (Se, 99.9999 \%). To ensure the removal of residual oxygen and potential contaminants, quartz ampoules were subjected to rigorous chemical cleaning and vacuum degassing prior to loading. The precursors were sealed in a quartz ampoule (inner diameter $\approx$ 10 mm; length $\approx$ 150 mm) using iodine (5 mg/cm$^3$) as the transport medium. Following evacuation to high vacuum, the ampoule was situated in a dual-zone horizontal furnace. Growth of high-quality crystals was governed by the precise optimization of the iodine concentration and the thermal gradient between the source and deposition zones. Specifically, the source was held at 960--980 $^\circ$C, while the growth zone temperature was ramped from 880 $^\circ$C to 900 $^\circ$C over a 100-hour period. These conditions were maintained for 300 hours to promote the development of large-scale crystals. A subsequent 100-hour programmed cooling phase reduced the source and growth regions to 200 $^\circ$C and 100 $^\circ$C, respectively, followed by ambient cooling to room temperature. The resulting crystals, typically measuring 5 $\times$ 5 $\times$ 0.1 mm$^3$, were rinsed with methanol to eliminate surface iodine. Initial compositional characterization was conducted via energy-dispersive X-ray spectroscopy (EDS) using a field-emission scanning electron microscope (FE-SEM, JEOL 7500).

\textit{\textbf{STM experiments: }} Samples were mounted to the holder utilizing silver conducting epoxy (EPO-TEK H20E) and cured for 20 minutes at 175 $^\circ$C; a cleaving rod was attached to the sample surface using an identical procedure. The crystals were cold-cleaved under ultra-high vacuum (UHV) at cryogenic temperatures (tens of Kelvin) and immediately transferred into the STM head. Measurements were performed with a customized Unisoku USM1300 microscope using homemade, chemically etched tungsten tips, which were UHV-annealed to a bright orange glow before use. Spin-polarized tips were fabricated in situ by scanning and applying bias pulses to the sample surface, facilitating the pickup of Co adatoms. The magnetic nature of these tips was subsequently verified by scanning the UHV-cleaved surface of Fe$_{1+x}$Te (see Supplementary Figure 4), which exhibits a known bicollinear antiferromagnetic (AFM) order. This calibration confirmed that the spin-polarized tips behaved as ferromagnetic probes, reversing their polarization in response to external magnetic field direction. Unless otherwise noted, STM data acquisition was conducted at a base temperature of approximately 4.8 K.\\

\textit{\textbf{STM analysis:}} To quantify the intensity of the 2 $\times$ 2 modulation peaks in the Fourier transform, topographs and density of states (DOS) maps were processed using the Lawler-Fujita drift-correction algorithm \cite{Lawler2010}. This procedure applies a transformation to the real-space data such that the atomic Bragg peaks are registered to single pixels at even integer coordinates in Fourier space, effectively neutralizing artifacts arising from thermal drift or piezoelectric nonlinearities. Because the 2 $\times$ 2 periodicities are commensurate with the lattice, they are similarly mapped to discrete pixels, allowing for precise intensity extraction. \\

\textit{\textbf{Phenomenological theory:}}

We construct a minimal Landau theory for nematicity in a $C_3$-symmetric altermagnetic metal. The low-energy electronic structure is modeled as a collinear altermagnet with two spin-split Fermi surfaces,
\[
H_0(\mathbf{k})=\xi(\mathbf{k})\sigma_0+\Delta_A f_A(\mathbf{k})\sigma_z .
\]
Here $\xi(\mathbf{k})$ is the spin-independent dispersion, $\sigma_0$ is the identity matrix in spin space, $\sigma_z=\pm 1$ labels the spin projection, and $\Delta_A$ is the altermagnetic order parameter. For the in-plane sector relevant to the three $C_3$-related electronic channels, we take the altermagnetic form factor to have the threefold structure
\[
f_A(\theta)\sim \cos 3\theta .
\]
This form factor satisfies
\[
f_A(\theta+2\pi/3)=f_A(\theta),
\qquad
f_A(\theta+\pi/3)=-f_A(\theta),
\]
so that a $C_3$ rotation leaves each spin-split sector invariant, while a $C_6$ rotation exchanges the two spin-split Fermi surfaces. We use this form factor as a minimal in-plane phenomenological description of the altermagnetic spin splitting.

We describe nematic distortions of the two spin-split Fermi surfaces by complex order parameters
\[
N_+=A_+e^{i\phi_+},\qquad
N_-=A_-e^{i\phi_-}.
\]
The phases are related to the physical nematic director angles by $\phi_\pm=2\alpha_\pm$, since nematicity is a quadrupolar order. Since the spin-up and spin-down bands are related by a 60 degree rotation, and since nematicity is $l=2$, $N_+$ relates to $N_-$ by
\[
N_- = e^{i2\pi/3}
\]
It is useful to define charge-like and spin-sensitive nematic combinations,
\[
\Phi=\frac{N_+ + N_-}{2},\qquad
\Psi=\frac{N_+ - N_-}{2}.
\]
The component $\Phi=|\Phi|e^{i\phi_\Phi}$ describes the spin-averaged, charge-like nematic response, while $\Psi=|\Psi|e^{i\phi_\Psi}$ describes the spin-sensitive nematic response. In spin-polarized STM, the latter is measured through its projection onto the tip polarization.

The most general quadratic free energy for $N_+$ and $N_-$, retaining the terms relevant to their relative phase, is
\[
F_2 =
r_N\left(|N_+|^2+|N_-|^2\right)
-J_R\left(N_+^*N_-+N_-^*N_+\right)
-J_I\Delta_A\, i\left(N_+^*N_- - N_-^*N_+\right).
\]
The $J_R$ term is a time-reversal-even inter-Fermi-surface locking term. The imaginary bilinear is time-reversal odd by itself; therefore the factor of $\Delta_A$ is required to make the free energy invariant in the altermagnetic state.

Using
\[
N_+=\Phi+\Psi,\qquad N_-=\Phi-\Psi,
\]
we obtain
\[
N_+^*N_-+N_-^*N_+=2\left(|\Phi|^2-|\Psi|^2\right),
\]
and
\[
i\left(N_+^*N_- - N_-^*N_+\right)
=
2i\left(\Psi^*\Phi-\Phi^*\Psi\right).
\]
The quadratic free energy can therefore be written as
\[
F_2 =
r_\Phi|\Phi|^2+r_\Psi|\Psi|^2
-2J_I\Delta_A\, i\left(\Psi^*\Phi-\Phi^*\Psi\right),
\]
where
\[
r_\Phi=2r_N-2J_R,\qquad
r_\Psi=2r_N+2J_R .
\]
The final term gives the relative phase locking,
\[
F_{\rm lock}
=
-4J_I\Delta_A|\Phi||\Psi|\sin(\phi_\Psi-\phi_\Phi).
\]
Thus, in the absence of lattice clock anisotropy, the altermagnetic coupling favors
\[
\phi_\Psi-\phi_\Phi=\pm \frac{\pi}{2},
\]
with the sign determined by $J_I\Delta_A$.

The $C_3$ lattice imposes additional clock terms on the nematic phases. For the charge-like component,
\[
F_\Phi^{\rm clock}
=
-\frac{\omega_\Phi}{2}\left(\Phi^3+\Phi^{*3}\right)
=
-\omega_\Phi|\Phi|^3\cos 3\phi_\Phi .
\]
For the spin-sensitive component, the corresponding clock term is allowed in the altermagnetic state as
\[
F_\Psi^{\rm clock}
=
-\frac{\omega_\Psi\Delta_A}{2}\left(\Psi^3+\Psi^{*3}\right)
=
-\omega_\Psi\Delta_A|\Psi|^3\cos 3\phi_\Psi .
\]
For a fixed altermagnetic domain and a suitable choice of phase convention, these clock terms pin the nematic phases to the three Potts values
\[
\phi_{\Phi,\Psi}=0,\frac{2\pi}{3},\frac{4\pi}{3}.
\]

The combination of the continuous altermagnetic locking term and the discrete $C_3$ clock anisotropy selects a neighboring Potts sector for the spin-sensitive component relative to the charge-like component. For example, choosing a charge nematic domain with $\phi_\Phi=0$, the allowed spin-sensitive phases are $\phi_\Psi=0,2\pi/3,4\pi/3$. The locking energy becomes
\[
F_{\rm lock}=-K\sin\phi_\Psi,\qquad
K=4J_I\Delta_A|\Phi||\Psi|.
\]
For $K>0$, the minimum occurs at $\phi_\Psi=2\pi/3$, while for $K<0$ it occurs at $\phi_\Psi=4\pi/3$. Thus, the charge-like and spin-sensitive nematic responses are shifted by one Potts sector. The two possible nonzero offsets are related by reversing the altermagnetic domain or the sign of the coupling.

Since $r_\Phi$ and $r_\Psi$ are generally different, the charge-like and spin-sensitive nematic components need not have identical onset temperatures. One component may condense first and induce the other through the altermagnetic locking term, or the two components may appear through separate transitions. The present low-temperature STM measurements establish the coexistence of the two components and their phase-shifted relation, while temperature-dependent measurements will be required to determine their onset sequence.\\

\noindent{\bf Acknowledgments}
The work was supported by the US Department of Energy grant number DE-SC0025005 (I.Z). Z.W. acknowledges the support of U.S. Department of Energy, Basic Energy Sciences Grant No. DE-FG02-99ER45747. F.M. acknowledges MUR funding within the FIS2 (n. 1236, 1-8-2023) Project no. FIS-2023-03652 (CUP D53C25001950001). Y.H. acknowledges the support from the National Research Foundation of Korea (NRF) grant funded by the Korea government (MSIT) (NRF-2022R1I1A1A01063507 and RS-2025-24742993)\\
\\
\noindent{\bf Data availability}\\
The data that support the findings of this study are available from the corresponding authors upon reasonable request.\\
\\
\noindent{\bf Code availability}\\
The code that supports the findings of the study is available from the corresponding authors upon reasonable request.\\
\\
\noindent{\bf Competing financial interests}\\
The authors declare no competing financial interests.\\

\bibliographystyle{custom-style.bst}


\newpage

\begin{figure} 
    \centering
    \renewcommand{\thefigure}{3 (cont.)}  
    \renewcommand{\figurename}{FIG.}
    \renewcommand{\baselinestretch}{1}
    \caption{\textbf{e}, Amplitudes of the 2$a_0$ Fourier transform peaks as a function of tiny magnetic field used to polarize the tip in different directions, showing an obvious switching of the $Q_2$ peak between positive fields (red background) and negative fields (blue background). \textbf{f}, $M_T$(\textbf{r}) map obtained by subtracting two topographs, both taken with a spin-up polarized tip, and corresponding Fourier transform (inset). \textbf{g}, Amplitudes of the 2$a_0$ Fourier transform peaks of the magnetic contrast map as a function of energy obtained by subtracting dI/dV maps with a spin-up and spin-down polarized tips. \textbf{h}, Amplitudes of the 2$a_0$ Fourier transform peaks as a function of energy in the charge map  obtained by adding dI/dV maps with a spin-up and spin-down polarized tips. STM setup conditions: (\textbf{a}) $V_{sample}$ = 50 mV, $I_{set}$ = 100 pA, spin-up tip; (\textbf{b}) $V_{sample}$ = 50 mV, $I_{set}$ = 100 pA, spin-down tip; \textbf{e} $V_{sample}$ = 50 mV, $I_{set}$ = 100 pA. \textbf{g, h} $V_{sample}$ = 50 mV, $V_{exc}$ = 5 mV (RMS), $I_{set}$ = 100 pA.}
\end{figure}

\begin{figure}
    \renewcommand{\thefigure}{4}  
    \renewcommand{\figurename}{FIG.}  
    \centering
    \includegraphics[width = \textwidth]{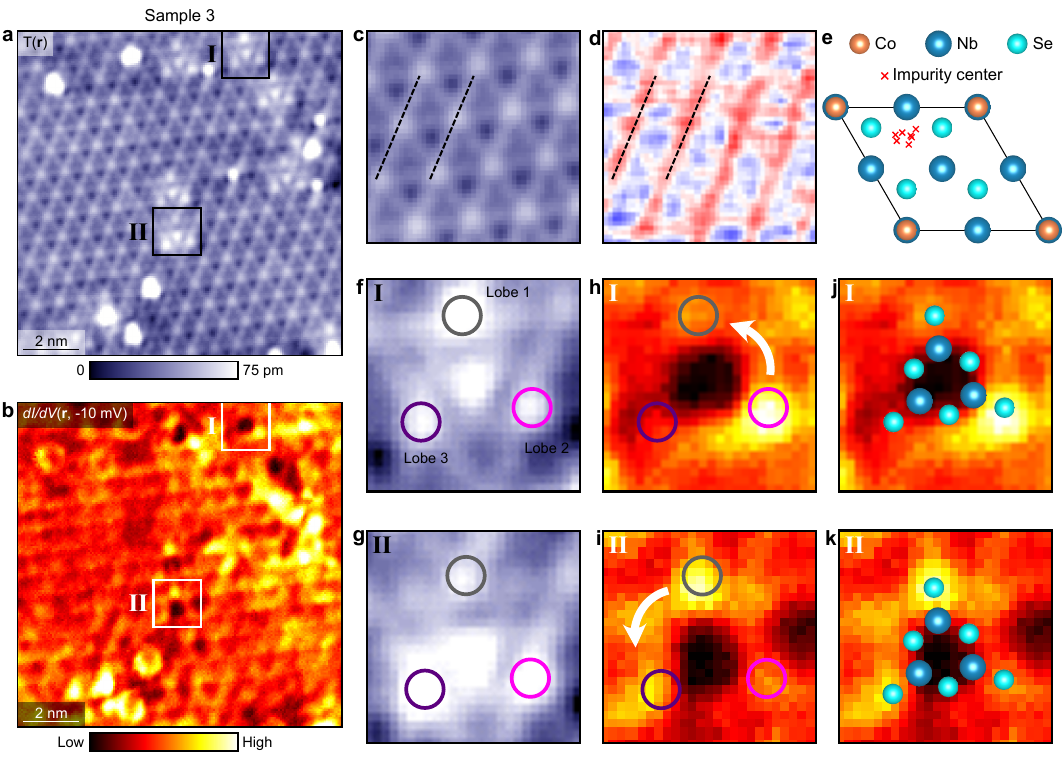}
    \renewcommand{\baselinestretch}{1}
    \caption{\textbf{Single atom impurities.} \textbf{a}, STM topograph of the Se termination with two examples of one type of impurity boxed. \textbf{b}, $dI$/$dV$ map taken over the same area in (\textbf{a}) with the same two impurities boxed. \textbf{c}, Zoom-in on an impurity-free area in (\textbf{a}). \textbf{d}, Magnetic contrast map of the area in (\textbf{c}) obtained by subtracting STM topographs with spin-up and spin-down polarized tips. \textbf{e}, Unit cell in the $ab$-plane showing the center of all 7 impurities of the same type within the field of view in (a) marked with red Xs. \textbf{f-g}, Zoomed-in STM topographs, and \textbf{(h,i)} corresponding d$I$/d$V$ maps of impurities I and II, respectively. White arrow denote direction from the brightest lobe to the second brightest lobe. \textbf{j-k}, Same image as in (\textbf{h-i}) with the atomic-structure overlaid. STM setup conditions: (\textbf{a}) $V_{sample}$ = 50 mV, $I_{set}$ = 200 pA; (\textbf{b}) $V_{sample}$ = -10 mV, $V_{exc}$ = 5 mV (RMS), $I_{set}$ = 200 pA. The magnetic field is 0 T.}
\label{fig:4}
\end{figure}

\begin{figure}
    \renewcommand{\thefigure}{5}  
    \renewcommand{\figurename}{FIG.}  
    \centering
    \includegraphics[width = \textwidth]{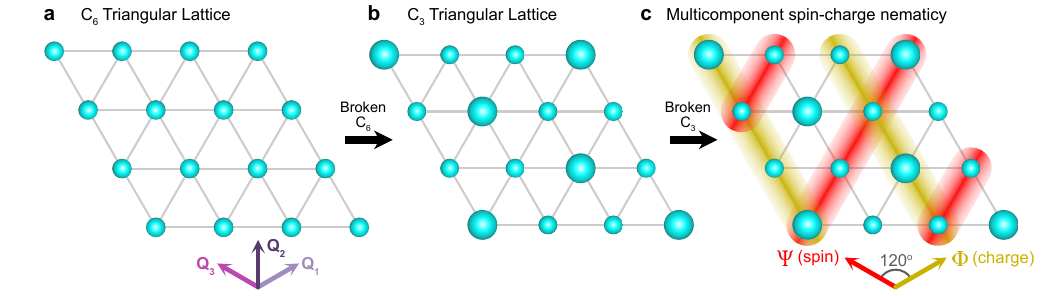}
    \renewcommand{\baselinestretch}{1}
    \caption{\textbf{Pedogogical cartoon of multicomponent nematicity.} \textbf{a}, Cartoon model of a topography of a $C_6$ symmetric triangular lattice, representing the Se termination before Co intercalation. \textbf{b}, Cartoon model of a topography of a $C_3$ symmetric triangular lattice, representing the Se termination after Co intercalation. \textbf{c}, Cartoon model of topography of a Co-intercalated Se termination with a multicomponent spin-charge nematicity, with the red stripes representing the spin component and yellow representing the charge component.}
\label{fig:5}
\end{figure}


\begin{figure}
    \renewcommand{\thefigure}{1}  
    \renewcommand{\figurename}{Extended Data FIG.}  
    \centering
    \includegraphics[width = \textwidth]{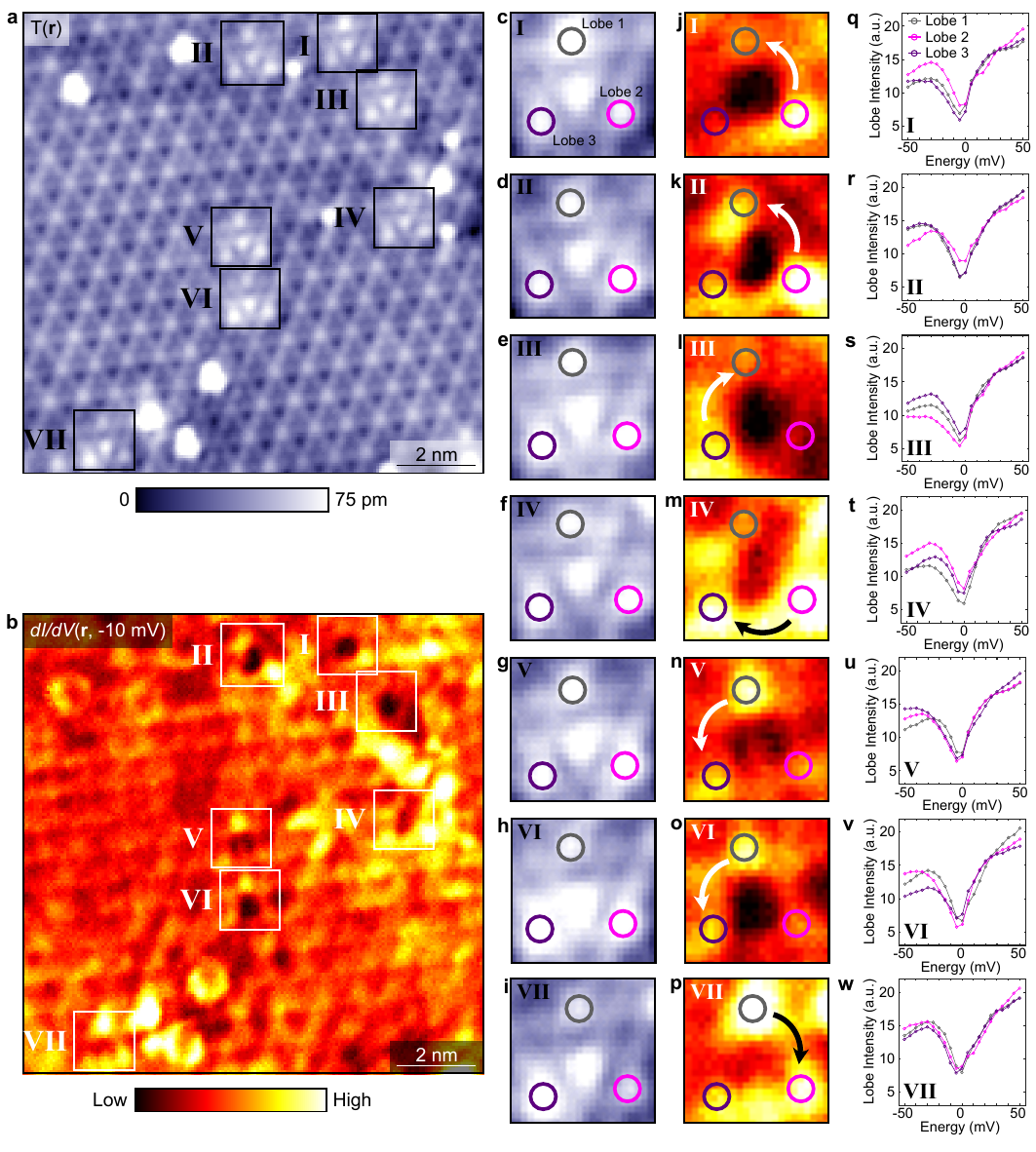}
    \renewcommand{\baselinestretch}{1}
    \caption{\textbf{Inequivalent symmetry of the impurities.} \textbf{a}, Topograph of the same area shown in Figure 4 with all 7 impurities numbered. \textbf{b}, $dI$/$dV$ map taken at -10 mV over the same area as in (\textbf{a}). \textbf{c-i}, Zoom-in of topograph over impurity I, II, III, etc., respectively, with lobe 1, 2, and 3 represented by a gray, pink, and purple circle, respectively. \textbf{j-p} Zoom-in of $dI$/$dV$ in (\textbf{b}) over impurity I, II, III, etc., respectively, with arrow pointing from brightest lobe to the second brightest lobe. \textbf{q-w}, Intensity of each lobe as a function of energy for impurity I, II, III, etc., respectively. STM setup conditions: (\textbf{a}) $V_{sample}$ = 50 mV, $I_{set}$ = 200 pA; (\textbf{b}) $V_{sample}$ = -10 mV, $V_{exc}$ = 5 mV (RMS), $I_{set}$ = 200 pA. The magnet was set to 0 T.}
\label{si:1}
\end{figure}


\begin{figure}
    \renewcommand{\thefigure}{2}  
    \renewcommand{\figurename}{Extended Data FIG.}  
    \centering
    \includegraphics[width = \textwidth]{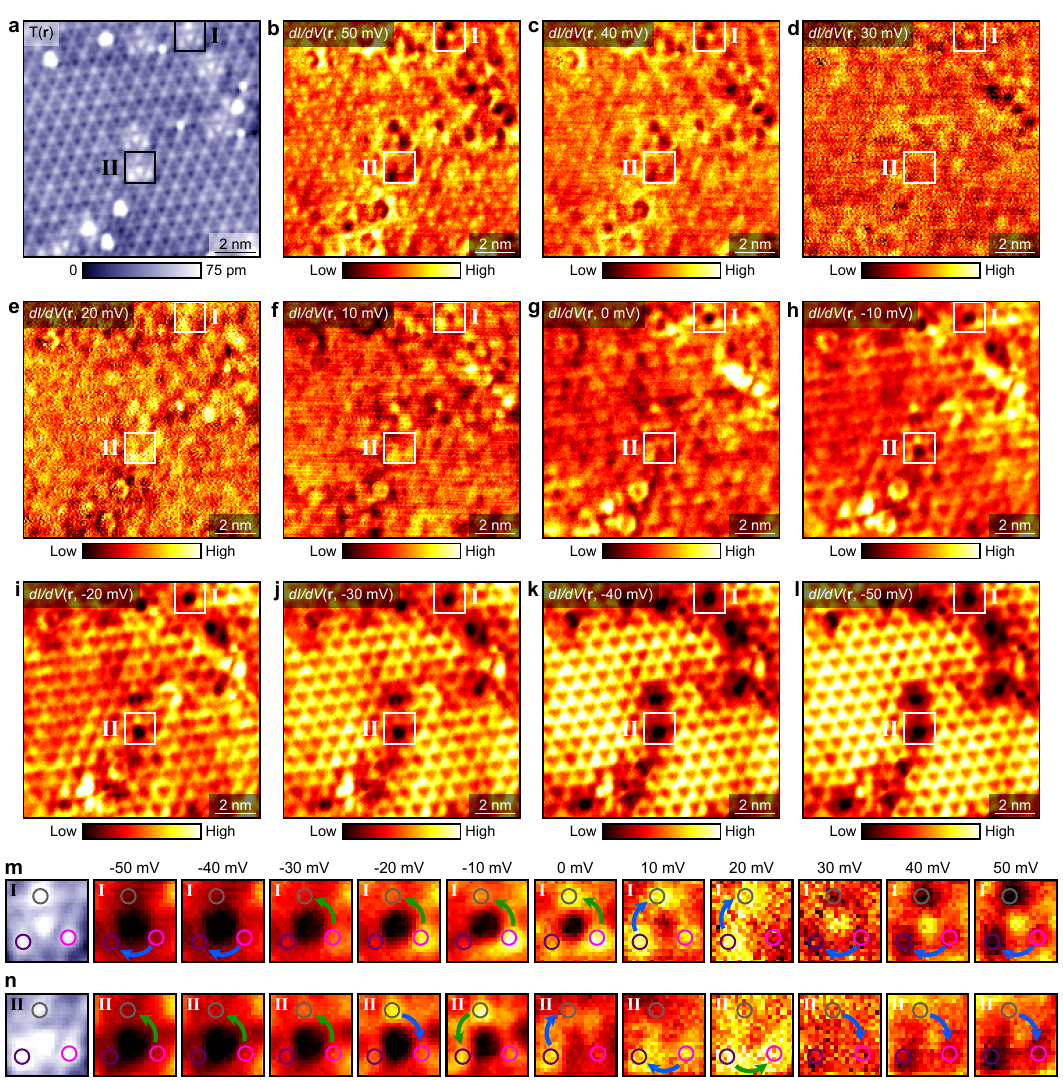}
    \renewcommand{\baselinestretch}{1}
    \caption{\textbf{Energy evolution of impurity states.} \textbf{a}, Topograph of the same area shown in Figure 4 with the two impurities numbered. \textbf{b-l} $dI$/$dV$ maps over the same area in \textbf{a} starting from 50 mV (\textbf{b}) to -50 mV (\textbf{l}) in steps of 10 mV. \textbf{m, n} Topographic zoom-in of impurity I and II with a zoom-in of their energy evolution, respectively. STM setup conditions: (\textbf{a}) $V_{sample}$ = 50 mV, $I_{set}$ = 200 pA; (\textbf{b-l}) $V_{sample}$ = 50 mV (\textbf{b}) to -50 mV (\textbf{l}) in steps of 10 mV, $V_{exc}$ = 5 mV (RMS), $I_{set}$ = 200 pA. The magnet was set to 0 T.}
\label{si:2}
\end{figure}

\begin{figure}
    \renewcommand{\thefigure}{3}  
    \renewcommand{\figurename}{Extended Data FIG.}  
    \centering
    \includegraphics[width = \textwidth]{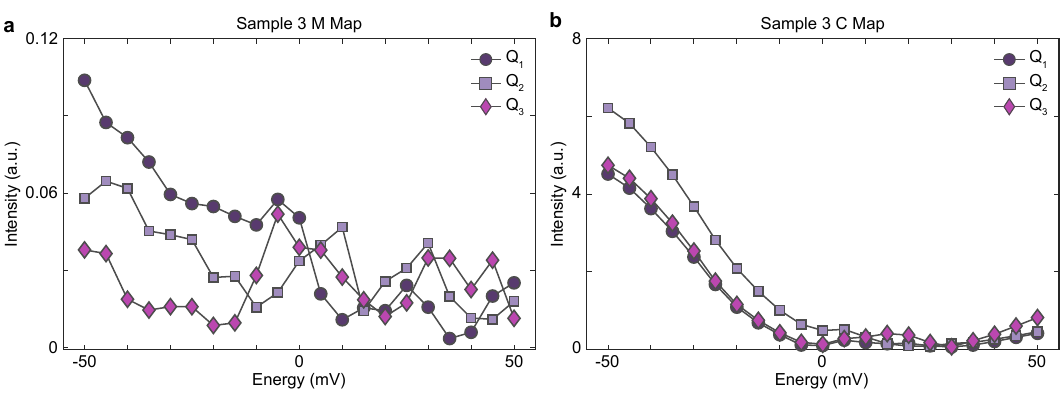}
    \renewcommand{\baselinestretch}{1}
    \caption{\textbf{Magnetic contrast and charge map analysis over sample 3}. \textbf{a}, Amplitudes of the 2$a_0$ Fourier transform peaks of the magnetic contrast map as a function of energy obtained by subtracting dI/dV maps with a spin-up and spin-down polarized tips. \textbf{h}, Amplitudes of the 2$a_0$ Fourier transform peaks as a function of energy in the charge map  obtained by adding dI/dV maps with a spin-up and spin-down polarized tips. STM setup conditions: (\textbf{a-b}) $V_{sample}$ = 50 mV, $V_{exc}$ = 5 mV (RMS), $I_{set}$ = 200 pA. The magnet was set to 0 T.}
\label{si:3}
\end{figure}

\begin{figure}
    \renewcommand{\thefigure}{4}  
    \renewcommand{\figurename}{Extended Data FIG.}  
    \centering
    \includegraphics[width = \textwidth]{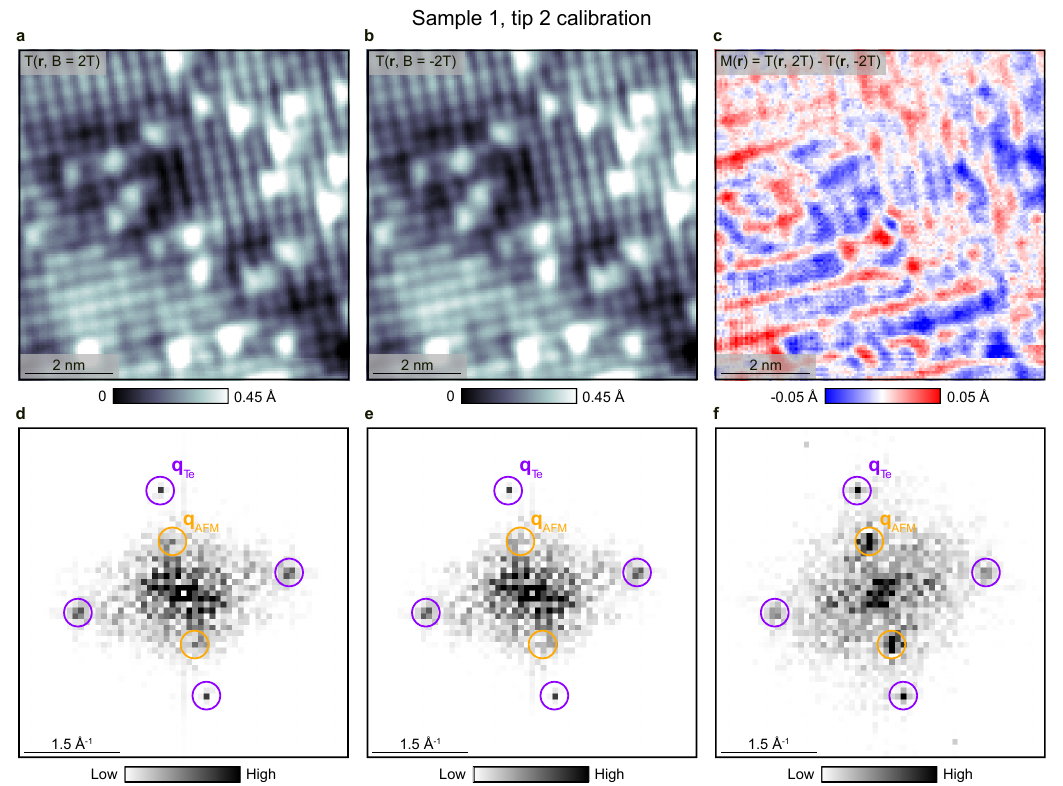}
    \renewcommand{\baselinestretch}{1}
    \caption{\textbf{STM sample 1, tip 2 calibration on FeTe}. \textbf{a}, Topography of a Te-terminated surface at +2 T with the same tip 2 used to scan sample 1. \textbf{b}, Topography of the same area as in (\textbf{a}) but at -2 T. \textbf{c}, Magnetic contrast map (M(\textbf{r}) map) taken by subtracting (\textbf{b}) from (\textbf{a}). \textbf{d-f}, Fourier transforms of (\textbf{a}), (\textbf{b}), and (\textbf{c}), respectively. The presence of the 2$a_0$ modulation seen by eye in the M(\textbf{r}) Map (\textbf{c}) and its Fourier transform (\textbf{f}) suggest a ferromagnetic spin-polarized tip. STM setup conditions: (\textbf{a-b}) $V_{sample}$ = 50 mV, $I_{set}$ = 100 pA.}
\label{si:4}
\end{figure}

\begin{figure}
    \renewcommand{\thefigure}{5}  
    \renewcommand{\figurename}{Extended Data FIG.}  
    \centering
    \includegraphics[width = \textwidth]{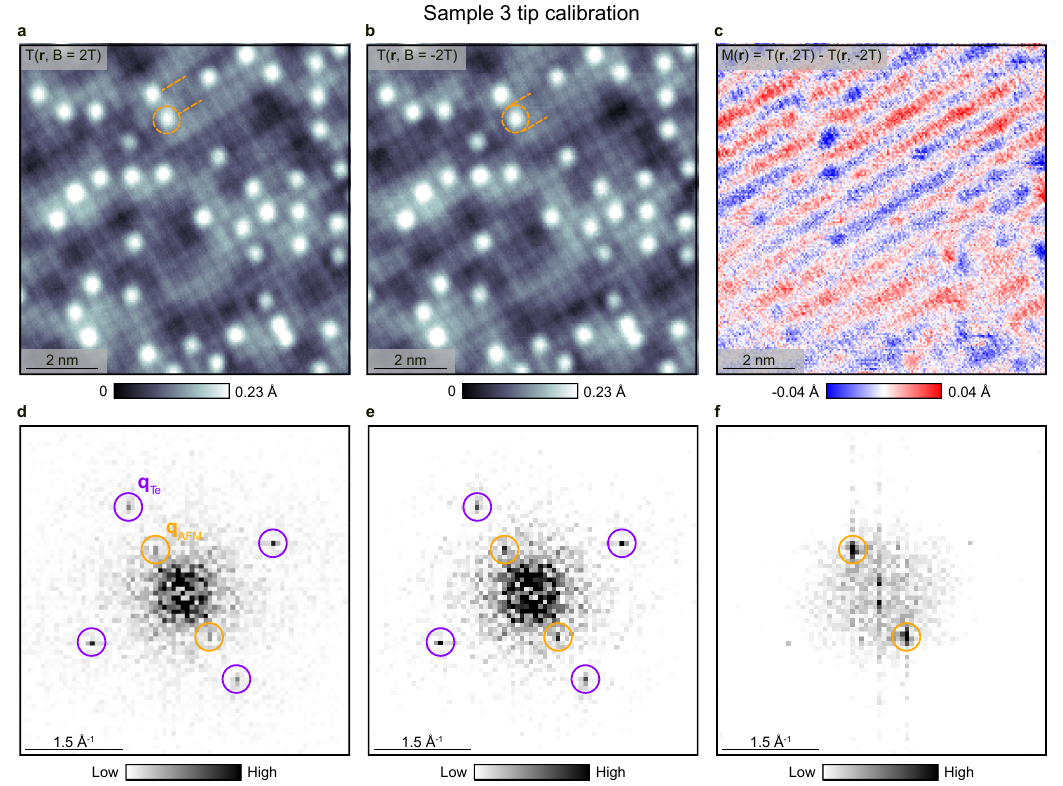}
    \renewcommand{\baselinestretch}{1}
    \caption{\textbf{STM sample 3 tip calibration on FeTe}. \textbf{a}, Topography of a Te-terminated surface at +2 T with the same tip used to scan sample 3. \textbf{b}, Topography of the same area as in (\textbf{a}) but at -2 T. \textbf{c}, Magnetic contrast map (M(\textbf{r}) map) taken by subtracting (\textbf{b}) from (\textbf{a}). \textbf{d-f}, Fourier transforms of (\textbf{a}), (\textbf{b}), and (\textbf{c}), respectively. The presence of the 2$a_0$ modulation seen by eye in the M(\textbf{r}) Map (\textbf{c}) and its Fourier transform (\textbf{f}) suggest a ferromagnetic spin-polarized tip. STM setup conditions: (\textbf{a-b}) $V_{sample}$ = 1 V, $I_{set}$ = 10 pA.}
\label{si:5}
\end{figure}

\end{document}